# Myths and Facts about a Career in Software Testing: A Comparison between Students' Beliefs and Professionals' Experience

Ronnie de Souza Santos[1,3], Luiz Fernando Capretz[2], Cleyton Magalhães[3], Rodrigo Souza[3]

[1] Shannon School of Business, Cape Breton University – Canada
[2] Department of Electrical & Computer Engineering, Western University – Canada
[3] Post-Baccalaureate program on Agile Testing, CESAR School – Brazil

Email: ronnie_desouza@cbu.ca, lcapretz@uwo.ca, cvcm@cesar.school, recs@cesar.school

**ABSTRACT**. Testing is an indispensable part of software development. However, a career in software testing is reported to be unpopular among students in computer science and related areas. This can potentially create a shortage of testers in the software industry in the future. The question is, whether the perception that undergraduate students have about software testing is accurate and whether it differs from the experience reported by those who work in testing activities in the software development industry. This investigation demonstrates that a career in software testing is more exciting and rewarding, as reported by professionals working in the field, than students may believe. Therefore, in order to guarantee a workforce focused on software quality, the academy and the software industry need to work together to better inform students about software testing and its essential role in software development.

## I. INTRODUCTION

Testing is an indispensable part of software development. It is responsible for verifying and validating systems and applications [1]. Currently, due to the agile nature of software development, testing is performed throughout the development process and the system's quality is continuously assessed [2][3]. Usually, testing processes take up about 40-50% of a project's time, directly impacting costs and deliveries.

The history of software engineering indicates that testing activities existed before the establishment of processes, practices, and models for software development [2][3][4]. However, in our current climate, a career focused on software testing appears to be undervalued by undergraduate students in software engineering and computer science programs [8]. A family of surveys recently conducted in different countries has demonstrated that undergraduate students see the work in software testing as monotonous, uninteresting, repetitive, stressful, and lacking in opportunities for coding tasks. In addition, students believe that a testing career is less relevant for software development than other career paths [5][6][7][8][9].

The unpopularity of software testing among students creates a major problem for software companies because the lack of skilled testing professionals negatively impacts the success of software projects [8]. However, as previous studies [5][6][7][8][9] have demonstrated, the perception students have about software testing is not based on experience. Instead, this perception is based mainly on particular beliefs about testing, which seem to reflect a lack of discussion about software testing in undergraduate courses. Therefore, it is in the best interests of the software industry to distinguish between myth and fact among the perspectives of computer science students and to begin designing strategies to increase the attractiveness of software testing careers.

In this study, we address this question by comparing the experiences of software professionals with the perceptions of undergraduate students about software testing careers. To do this, we applied the same survey questionnaire used by the previous studies [5][6][7][8][9] to assess the perceptions of experienced software professionals about a career in software testing. Our findings demonstrate that a career in software testing is very different from what students perceive, and most of the beliefs regarding testing tasks are more myth than fact.

**II. TESTING CAREER: SURVEYS OF STUDENTS**

Previous surveys analyzed the perceptions of 648 students from software engineering and related fields about software testing. These students were from seven different countries: 132 from the United Arab Emirates (UAE) [7], 99 from China [5], 92 from Brazil [8], 88 from Pakistan [9], 85 from Canada [5], 82 from Malaysia [6], and 70 from India [5]. They were asked about issues related to careers in software testing: 1) Motivators/PROs for taking up a testing career; 2) De-motivators/CONs for taking up a testing career; 3) The chances of them taking up a testing career.

As demonstrated in Table 1, the number of students who declare having no interest in a career in software testing is always much higher than the number of students opting to follow this path in the software industry. We illustrate some of the students' perspectives in Table 2.

**Table 1. Chances of taking up a testing career [5][6][7][8][9]**

| Answer | Brazil | Canada | China | UAE | India | Malaysia | Pakistan |
|---|---|---|---|---|---|---|---|
| Certainly No | 11% | 31% | 24% | 7% | 14% | 1% | 15% |
| No | 16% | 27% | 0% | 17% | 31% | 7% | 16% |
| Maybe | 46% | 33% | 74% | 53% | 47% | 52% | 47% |
| Yes | 14% | 7% | 2% | 18% | 7% | 34% | 14% |
| Certainly Yes | 13% | 2% | 0% | 5% | 0% | 6% | 8% |

**Table 2. Students' reasons for choosing a career in testing [8]**

| After you graduate, would you consider a career in software testing? | |
|---|---|
| **Answer** | **Justification (quotations)** |
| **Certainly No / No** | "I am not enthusiastic about software testing. I will rather be coding new features". <br><br> "Certainly, never caught my attention". <br><br> "I want to create games in the future, not this thing." <br><br> "Testing is not really to my taste. I like challenges and working with new people, new problems…" |
| **Certainly Yes / Yes** | "Since I am working with this lately, I don't really see me working with something else [after graduation]. This is what I like". <br><br> "I liked it since I learned about it in a lecture." <br><br> "Testing is an area that caught my attention, it will give the opportunity to learn a lot". |

There are two main reasons identified by students who prefer not to work with software testing after graduation (Certainly No/No responses). First, some of them have already developed interests in other areas of software development and expect to work in those areas in the future. Second, some have an outdated perception of the career and see testing jobs as monotonous, repetitive, and less technical than other careers. In addition, many students seem unaware of the importance of software testing for the software development process, e.g., they cannot even imagine the connection between testing and other software development activities.

On the other hand, those who answered yes to the possibility of a career in software testing (Certainly Yes/Yes responses) seem to have had previous contact with the area, either they have taken an internship that included testing or know someone who works as a software tester. In addition, there are students who attended courses or lectures focused on software testing, and therefore, they understood the role of software testing in the industry and developed a positive attitude toward this career.

Finally, there are students who do not have software testing as their main interest, but they are open to following a career in this area, depending on several factors, such as compensation, benefits, opportunities for learning, and self-development opportunities. These students mostly answered *maybe* on the surveys, and their answers demonstrate that the lack of knowledge about *what in fact software testing is* seems to determine whether they are willing to follow this career or not.

**III. METHOD**

Since the scenario described in Section II seems to be more related to an impression or a belief that students have about a testing career and less to an understanding of the role of testing in the software industry, we *replicated* the survey with practitioners and assessed their experience regarding software testing careers.

A replication is a conscious and systematic repetition of a study, which is mainly used to generalize results in different populations by challenging or confirming previous findings [10]. In this study, we followed the same protocol used by the survey with students [5][6][7][8][9], apart from a few adaptations in the questionnaire to align with practitioners' roles, as described below.

*A. INSTRUMENT*

We changed the wording of the questions to refer to practitioners instead of software engineering students. Using this questionnaire, we wanted to understand how practitioners who are currently working in the field see the software testing career and compare this against the beliefs of students, i.e., those who have little or no practical experience with software testing activities. Table 3 presents the adapted questionnaire.

**Table 3. Questionnaire**

| According to the SWEBOK, software testing is defined as the dynamic process of verifying and validating software under development to attest that it works as expected and possesses all the planned features and behaviors. Based on this definition and on your previous knowledge/experience with software testing in the software industry, please answer the following questions. | |
|---|---|
| **Topic** | **Questions** |
| **Background** | 1. What is your current role in software development? |
| | 2. How many years of experience do you have in software development? |
| **Software Testing Career** | 3. Cite up to three Motivators/PROs for working in a testing career? |
| | 4. Cite up to three De-motivators/CONs for working in a testing career? |
| | 5. Will you continue working with software testing, or if you are not in a testing career, would you change your career to pursue one? <br> ( ) Certainly No <br> ( ) No <br> ( ) Maybe <br> ( ) Yes <br> ( ) Certainly Yes |

*B. SAMPLING AND DATA COLLECTION*

We followed recommendations to treat software professionals as a hidden population to mitigate the difficulty of reaching individuals to answer the survey. In this process, we applied two different techniques to send our questionnaire to practitioners: a) convenience sampling to select participants from our extensive network of professionals based on their availability; b) snowballing sampling by asking participants from the survey to forward the questionnaire to other professionals, such as colleagues and coworkers [11].

*C. DATA ANALYSIS*

The data collected in this study are mainly quantitative, i.e., composed of demographic/background questions and items (e.g., motivators/demotivators) that can be easily converted into quantitative data. Therefore, we applied descriptive statistics to analyze and summarize the information emerging from our data set. Descriptive statistics allowed us to present the distribution and frequency of participants' answers regarding their experience with software testing [12].

*D. ETHICS*

To guarantee participants' anonymity, no personal information about the participants was collected in this study (e.g., name, e-mail, or employer). In addition, participants were informed about the use of the collected data for scientific purposes.

## IV. FINDINGS

We obtained 63 answers from practitioners in the software industry. Our sample has an average experience of 8.7 years working with software development, with the most experienced professional in the sample having worked in the software industry for 25 years, and the least experienced for two years. Our sample is composed not only of professionals who work directly with software testing; it also includes programmers, software project managers, software architects, and designers. Figure 1 shows the distribution of our sample. We decided to include other professionals in addition to software testers because: a) in agile software development, testing activities are widely distributed throughout the whole process and therefore influence other professionals [2][3]; and b) comparing the perspective of other professionals with students provides a better understanding of myths and realities of software testing.

Figure 1. Distribution of professionals in the sample

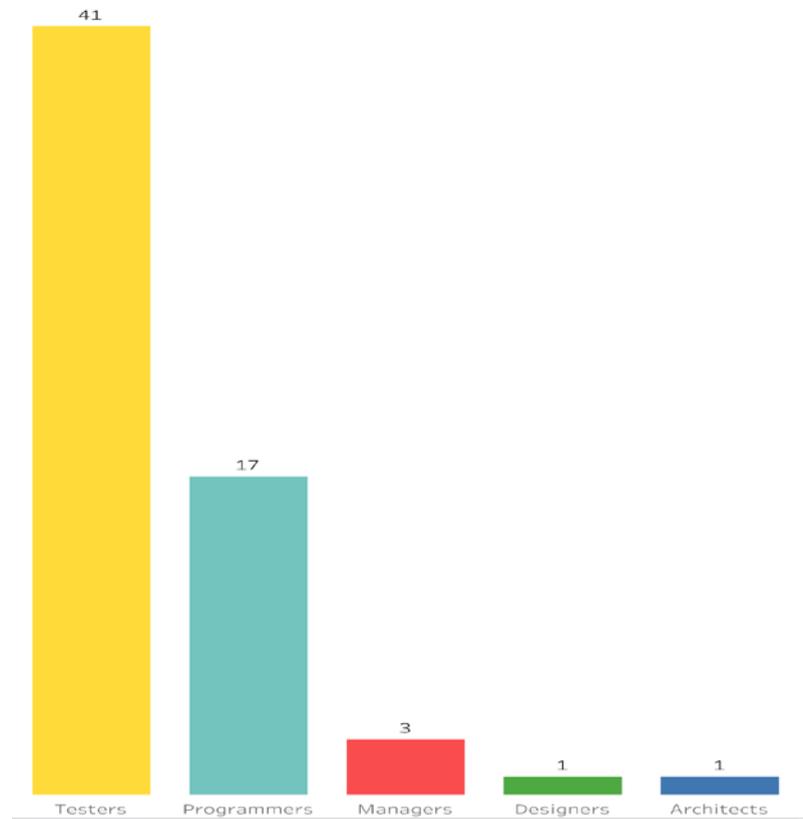

Our study identified several benefits reported by practitioners in their work in software testing, which are the main motivators for testing professionals. At the same time, they show how other professionals working in software development perceive testing activities. Findings demonstrate that 24% of our sample perceive software testing as a challenging activity, as testing professionals are usually the main ones responsible for guaranteeing the quality of software products and the success of project deliveries. In addition, 24% of professionals identified the number of job opportunities as one of the main positive factors of a career in software testing, given the high demand for these skills. Furthermore, 22% of the sample respondents highlighted that software testing involves a variety of tasks since testing professionals are required to plan and execute several scenarios that can be easily overlooked in other steps of software development. This variety also includes mastering several techniques, tools, and technologies. About 19% of the sample pointed out that a software testing career is well-recognized in the software industry. In addition, 17% of respondents highlighted that programming is a great deal of work for software testing professionals because automation is one important skill now required for these professionals. Other advantages of pursuing a testing career, according to our sample of professionals, include competitive salary (16%), continuous learning opportunities (14%), teamwork dynamics (5%), flexible jobs (5%), and stability (1%). Figure 2 illustrates these findings.

Regarding drawbacks that can demotivate professionals in a software testing career, our results indicate that 27% of professionals believe that a job in software testing pays less than other jobs in software development and that this is the main factor that drives people to other careers. About 24% of professionals perceived that a career in software testing does not offer visibility and recognition compatible with the importance of the work to the success of software projects. Together, 48% of the sample highlighted a lack of visibility or recognition as the reason the career might be overlooked by those entering the field. Comparing benefits and limitations so far, salary is a motivator/demotivator that varies in our sample; however, recognition/visibility tends to be more of a problem than an advantage in the career. Our results demonstrated several other demotivators perceived by professionals. These are deficiency in management processes, which usually has negative effects on test planning and execution (17%); repetitiveness (13%) and simplicity (8%), usually associated with testing environments with little or no opportunities for automation; increased workload (8%) resulting from poor management; high complexity (8%), unpredictability (6%), depending on the type of system and characteristics of the clients and users; and finally, lack of support regarding appropriate tools and technologies to perform the testing tasks (5%).

**Figure 2. Benefits of working with Software Testing**

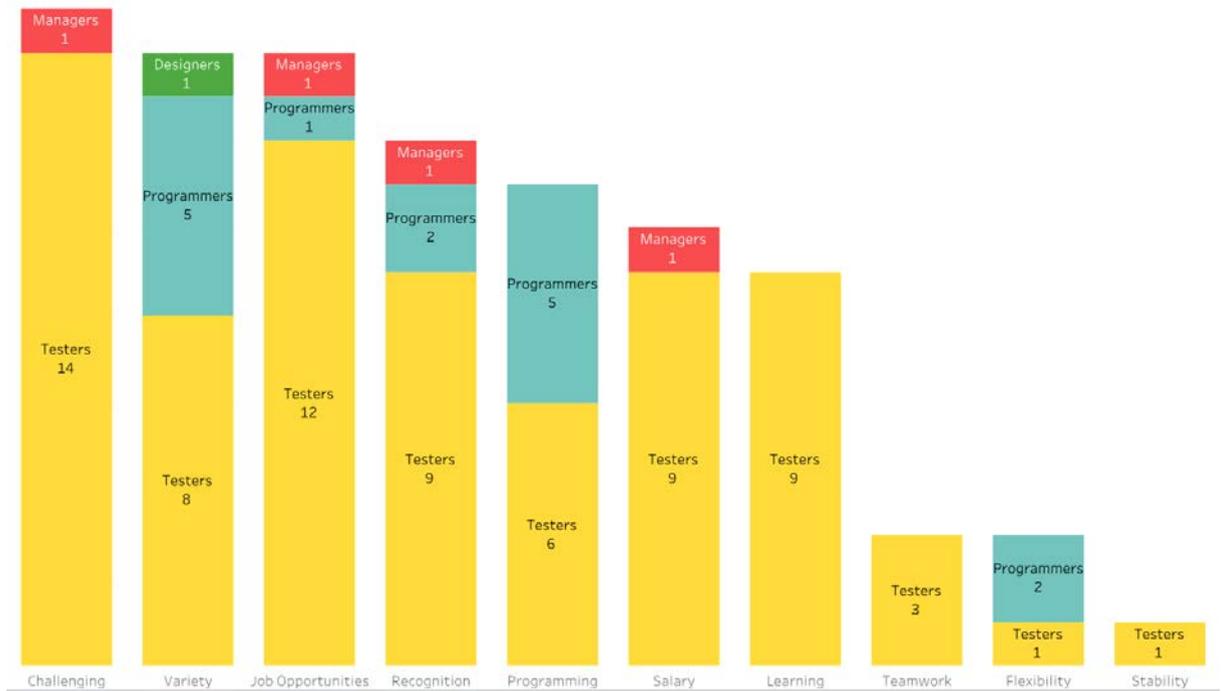

**Figure 3. Drawbacks of working with Software Testing**

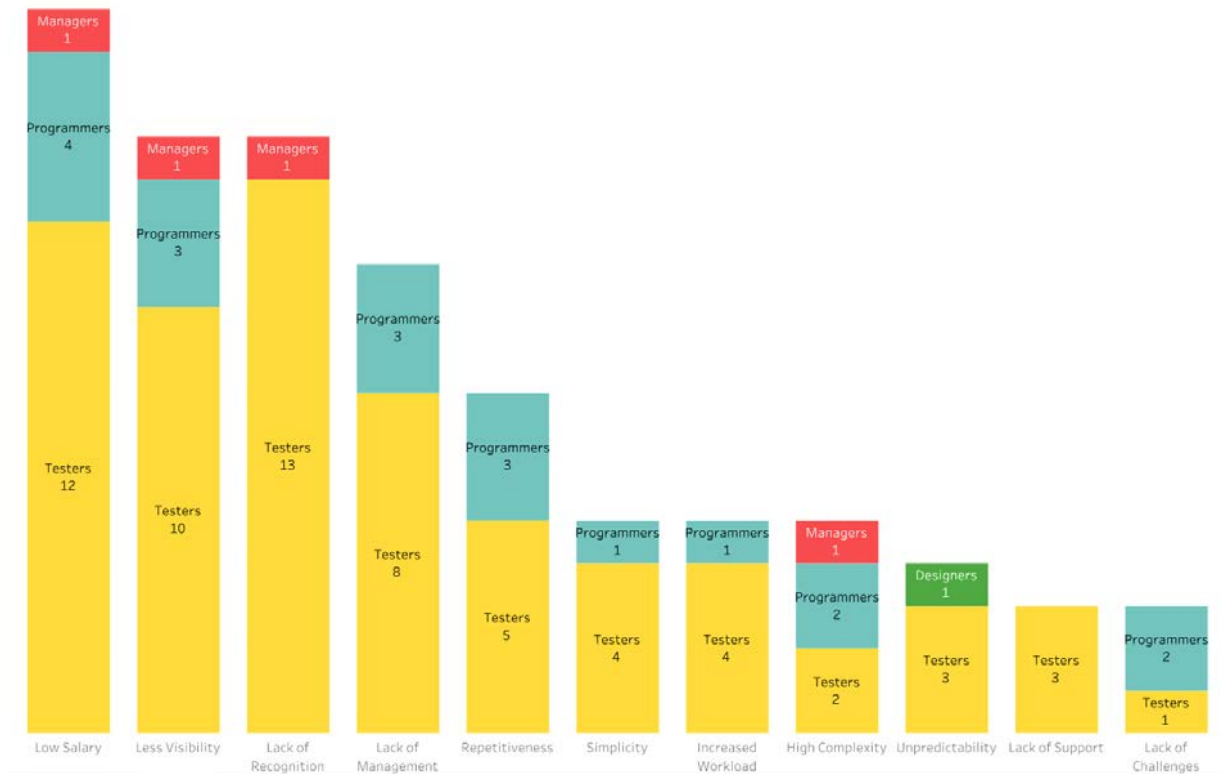

Looking specifically at software testing professionals, we observed that the main career motivators reported by professionals are the chance to work with challenging tasks (22%), the increasing number of job opportunities available (19%), the learning opportunities provided by this area of software development (14%), and the variety of activities that professionals can perform (13%). The main career drawbacks reported are lack of recognition (21%), less competitive salaries (19%), less visibility (16%), and lack of good management (13%). These responses reveal that the difficulties of software testing careers are not related to the characteristics of the tasks themselves. On the contrary, testing professionals are motivated by their tasks; however, they are unsatisfied with how they are treated in the software industry.

To answer the main question that motivated the present study, we asked our sample of participants how they felt about continuing to work with software testing (if they are already testing professionals) or whether they would consider switching careers and starting to work in software testing (if they have other roles in the software development process). In summary, software testers are very positive about their career, with 93% of the sample affirming that they intend to keep working in this area (answering certainly yes and yes), while 7% apparently would consider changing careers (answering maybe). According to our analysis, those who are considering other careers may be basing their decision on the perspective of receiving a better salary by working in other areas, but their views seem to be mostly due to problems with a lack of good management and a lack of visibility of their work. Regarding other professionals who participated in the study, i.e., programmers, managers, designers, and software architects, those who reported that they would consider a career in software testing are motivated by the possibility of working with a variety of tasks that support the quality of software projects. In particular, programmers who would consider a career in testing identify test automation as one of the main motivators. Figure 4 illustrates this result.

**Figure 4. Working with Software Testing**

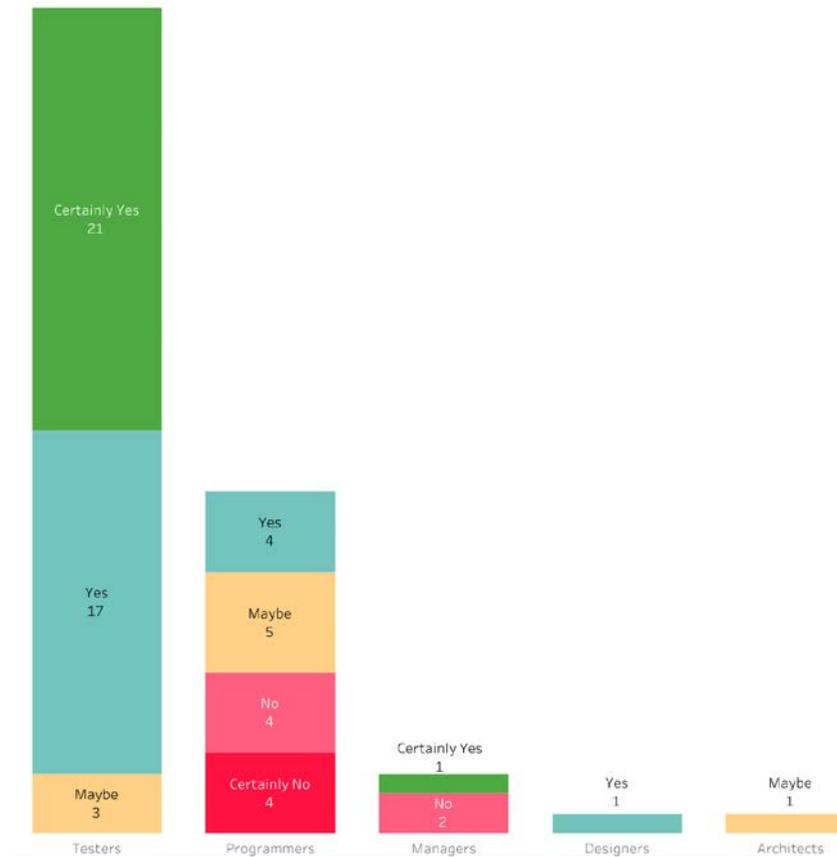

## V. DISCUSSIONS

### A. PURSUING A CAREER IN SOFTWARE TESTING

Our study presents a snapshot of how professionals currently working in the software industry perceive software testing. Testing is an essential activity in software development involving a variety of tasks that challenge professionals every day and provide them with the possibility of learning and developing several skills, both soft skills, e.g., teamwork dynamics, and hard skills, e.g., programming to enable test automation. One important result presented in this research is that these positive characteristics of software testing are not only perceived by those who work directly with it (i.e., testers and QAs) but also by other professionals involved in the software development process, including programmers, managers, and designers.

Those who are pursuing a career in software testing are mainly motivated by the challenges that this work provides and by the availability of job opportunities that allow them to select the best environments to work in. Regarding the work environment, those professionals following a career in software testing tend to prefer flexible environments that allow them to work with dynamic teams and provide them with the opportunity to work with a variety of scenarios. Speaking of variety, test automation now plays an important role in the career of software testing professionals.

However, our results revealed one issue that could affect the software industry greatly. Although testing professionals report the increasing number of job opportunities available as a motivator for choosing this career, this factor might also be revealing a problem that the software industry will soon face, namely a shortage of qualified professionals to perform quality activities in software projects. At least two factors are likely contributing to this shortage. First, undergraduate students see software testing as a backup career rather than their first choice, which suggests this shortage will only increase. Second, testing professionals are struggling with bad management, increased workload, and lack of recognition, which might influence them to find alternative careers and migrate to other areas of software development. Software companies cannot afford to lose more qualified software testing professionals; therefore, these issues must be urgently addressed.

### B. BRINGING TALENTS INTO SOFTWARE TESTING

Another alternative to bringing more talented professionals to work in testing in the software industry is to change the perception that students have about the career and therefore increase the popularity of this area in software engineering and related courses. One important discussion about this issue is that students believe that software testing is mainly a monotonous, repetitive, and non-technical activity. However, as we have shown, this view is not necessarily accurate. Our results demonstrate that software testing encompasses a variety of dynamic and challenging activities, which vary from manual practices to coding scripts to automating tests. This provides professionals with opportunities to learn and improve their skills continuously, which according to recent studies, are essential motivators for software testing professionals [13][14].

Previous studies have demonstrated that students who had contact with testing activities through courses or internships have a more positive attitude toward pursuing a career in software testing [8]. Therefore, one of the keys to recruiting more talent to work in this area in the software industry is to increase the number of courses that focus on software quality at the undergraduate level. In particular, courses in testing automation are expected to increase the popularity of the area among students, helping them to form a new perception of the career.

## VI. CONCLUSIONS

The present study focused on comparing myths and facts about a career in software testing. Our results demonstrate that one common myth is that those working with software testing are destined to deal with repetitive and monotonous tasks in software development. This is an outdated perception of the career that is negatively affecting present-day undergraduate students. The reality is that a career in software testing can be dynamic and exciting, according to professionals who work with software development. However, as with any other job, there is room for improving the software development environment and implementing changes that will benefit testing professionals who might be demotivated with their careers; these changes mostly depend on managerial and organizational improvements. This study is a wake-up call for both academia and industry. On the one hand, academia needs to develop strategies to guarantee that students have an accurate perspective of the work in the software industry, which will help to reduce the unpopularity of some careers, including software testing. On the other hand, the software industry needs to be aware of the impact that a shortage of testing professionals might have on software development and therefore create strategies to improve the visibility of the career.

## AUTHORS

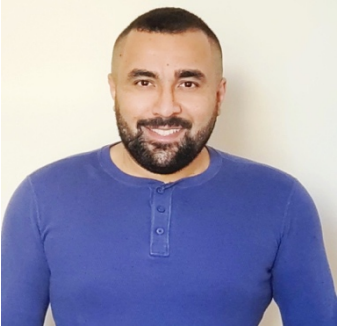

**Ronnie de Souza Santos** is an assistant professor at Cape Breton University in Canada. He received his Ph.D. in computer science from the Federal University of Pernambuco (Brazil) and completed 2 years of postdoctoral fellowship in software engineering at Dalhousie University in Canada. His research interests are equity, diversity, and inclusion in software engineering and software development practices. He is an adjunct professor at CESAR School (Brazil).

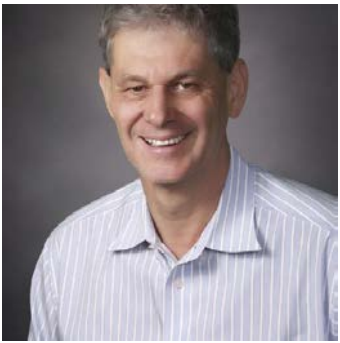

**Luiz Fernando Capretz** is a professor of software engineering and the director of the software engineering program at Western University in Canada. He received his B.Sc. in computer science from UNICAMP (Brazil), M.Sc. in applied computer science from INPE (Brazil), and Ph.D. in computing science from Newcastle University (UK). He has held visiting professorships at Yale-NUS MCS Program in Singapore, in the Department of Information and Computer Sciences at Universiti Teknologi PETRONAS in Malaysia, and in the Department of Computer Science at New York University and University of Sharjah - both in the UAE. His current interests lies in software testing and human factors in software engineering.

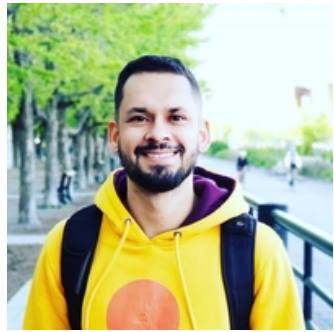

**Cleyton Magalhaes** is a professor in the Post-Baccalaureate program on Agile Testing at CESAR School in Brazil. He received his Ph.D. in computer science from the Federal University of Pernambuco (Brazil). He is a software QA at the Recife Center for Advanced Studies and Systems (CESAR).

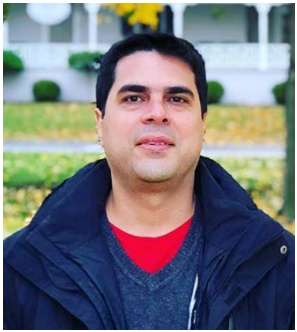

**Rodrigo Souza** is a former student in the Post-Baccalaureate program on Agile Testing at CESAR School in Brazil. He has been working with software testing since 2005. He is a software QA at Hearst.